\documentstyle[11pt,epsfig]{article}
\date{}
\begin{document}
\title{{\bf Scalar field quantum cosmology: a Schr\"{o}dinger picture}}
\author{Babak Vakili\thanks{%
email: b-vakili@iauc.ac.ir}\\\\
{\small {\it Department of Physics, Chalous Branch, Islamic Azad
University (IAU), P.O. Box 46615-397, Chalous, Iran}}} \maketitle

\begin{abstract}
We study the classical and quantum models of a scalar field
Friedmann-Robertson-Walker (FRW) cosmology with an eye to the
issue of time problem in quantum cosmology. We introduce a
canonical transformation on the scalar field sector of the action
such that the momentum conjugate to the new canonical variable
appears linearly in the transformed Hamiltonian. Using this
canonical transformation, we show that, it may lead to the
identification of a time parameter for the corresponding dynamical
system. In the cases of flat, closed and open FRW universes the
classical cosmological solutions are obtained in terms of the
introduced time parameter. Moreover, this formalism gives rise to
a Schr\"{o}dinger-Wheeler-DeWitt equation for the
quantum-mechanical description of the model under consideration,
the eigenfunctions of which can be used to construct the wave
function of the universe. We use the resulting wave functions in
order to investigate the possible corrections to the classical
cosmologies due to quantum effects by means of the many-worlds and
ontological interpretation of quantum cosmology.
\vspace{5mm}\noindent\vspace{0.8mm}\newline Keywords:
Scalar field cosmology, Quantum cosmology, Problem of time
\end{abstract}
\section{Introduction}
Since $1967$ when canonical quantum theory of gravity was first
introduced by DeWitt \cite{De}, many efforts have been made in
this area and the corresponding results have been followed by a
number of works, the main motivations of which lie in the results
of the unification theories and cosmology \cite{Vil}. The main
object of such a theory, i.e., the quantum state of the
gravitational system, is then described by a wave function, a
functional (on superspace) of the $3$-geometry and matter fields
presented in the theory, satisfying the Wheeler-DeWitt (WDW)
equation \cite{Kief}. One of the most important features related
to the WDW wave function is a problem which is known as the
"problem of time". Indeed, unlike the case of usual quantum
theory, the wave function in quantum gravity is independent of
time, which is a reflect to the fact that general relativity is a
parameterized theory in the sense that its (Einstein-Hilbert)
action is invariant under time reparameterization. Because the
existence of this problem, the canonical formulation of general
relativity leads to a constrained system, and its Hamiltonian is a
superposition of some constraints, the so-called Hamiltonian and
momentum constraints. A possible way to overcome this problem is
that one first solves the equation of constraint to obtain a set
of genuine canonical variables with which one can construct a
reduced Hamiltonian. In this kind of time reparameterization, the
equations of motion are obtained from the reduced physical
Hamiltonian and describe the evolution of the system with respect
to the selected time parameter \cite{Vak}.

The problem of time was first addressed in \cite{De} by DeWitt
himself. However, he argued that the problem of time should not be
considered as a hindrance in the sense that the theory itself must
include a suitable well-defined time in terms of its geometry or
matter fields. In this scheme time is identified with one of the
characters of the geometry, usually the scale factors (in
cosmological models) of the geometry and is referred to as the
intrinsic time, or with the momenta conjugate to the scale
factors, or even with a scalar character of matter fields coupled
to gravity in any specific model, known as the extrinsic time.
Identification of time with one of the dynamical variables depends
on the method we use to deal with theses constraints. Different
approaches arising from these methods have been investigated in
detail in \cite{Is} and \cite{time}.

In this paper, we first deal with a FRW cosmology with a scalar
field minimally coupled to the gravity. The phase-space variables
of such a model turn out to correspond to the scale factor $a(t)$
of the cosmological model and a scalar field $\phi(t)$ with which
the action of the model is augmented. We then introduce a
canonical transformation on the scalar field sector of the action
such that in terms of the new canonical variables, the transformed
Hamiltonian contains a linear momentum. This procedure causes the
Hamiltonian takes the form of a Schr\"{o}dinger one in which the
variable its conjugate momentum appears linearly in the
Hamiltonian plays the role of a time parameter. Moreover, this
formalism gives rise to a Schr\"{o}dinger-Wheeler-DeWitt (SWD)
equation for the quantum-mechanical description of the model under
consideration, the eigenfunctions of which can be used to
construct the wave function of the universe. In the cases of flat,
closed and open FRW models we present the classical and quantum
cosmological solutions in terms of the introduced time variable
and investigate the behavior of the corresponding universe in each
case separately.

\section{The classical model}
In this section we consider a FRW cosmology with a scalar field
with which the action of the model is augmented. In a
quasi-spherical polar coordinate the geometry of such a space-time
is described by the metric
\begin{equation}\label{A}
ds^2=-N^2(t)dt^2+a^2(t)\left[\frac{dr^2}{1-kr^2}+r^2\left(d\vartheta^2+\sin^2\vartheta
d\varphi\right)\right],\end{equation}where $N(t)$ is the lapse
function, $a(t)$ the scale factor and $k=1$, $0$ and $-1$
corresponds to the closed, flat and open universe respectively.
Since our goal is to study a procedure in which how a variable may
play the role of a time parameter, we do not include any matter
contribution in the action. Let us start from the action (we work
in units where $c=\hbar=16\pi G=1$)
\begin{equation}\label{B}
{\cal S}=\int d^4x\sqrt{-g}\left[R-\frac{1}{2}g^{\mu
\nu}\partial_{\mu}\phi\partial_{\nu}\phi-V(\phi)\right],\end{equation}where
$g$ is the determinant of the metric tensor $g_{\mu \nu}$, $R$ is
the Ricci scalar corresponding to $g_{\mu \nu}$ and $V(\phi)$ is
the potential function for the scalar field $\phi(t)$.  By
substituting (\ref{A}) into (\ref{B}) and integration over spatial
dimensions, we are led to a point-like Lagrangian in the
minisuperspace $\{a,\phi\}$ as
\begin{equation}\label{C}
{\cal L}=-3a\dot{a}^2+\frac{1}{2}
a^3\dot{\phi}^2+3ka-a^3V(\phi),\end{equation}in which we have set
$N=1$ so that the time parameter $t$ is the usual cosmic time. The
momenta conjugate to each of the above variables can be obtained
from their definition as
\begin{equation}\label{D}
P_a=\frac{\partial{\cal
L}}{\partial\dot{a}}=-6a\dot{a},\hspace{0.5cm}P_{\phi}=\frac{\partial{\cal
L}}{\partial\dot{\phi}}=a^3\dot{\phi}.\end{equation} In terms of
these conjugate momenta the canonical Hamiltonian, which is
constrained to vanish, is given by
\begin{equation}\label{E}
H=-\frac{P_a^2}{12
a}+\frac{P_{\phi}^2}{2a^3}-3ka+a^3V(\phi).\end{equation} Now, we
consider the following canonical transformation
$(\phi,P_{\phi})\rightarrow (T,P_T)$ on the scalar field sector of
the action \cite{Faraj}
\begin{equation}\label{F}
T=\frac{\phi}{P_{\phi}},\hspace{0.5cm}P_T=\frac{P_{\phi}^2}{2}.\end{equation}Under
this transformation Hamiltonian (\ref{E}) takes the form
\begin{equation}\label{G}
{\cal
H}=-\frac{P_a^2}{12a}+\frac{P_T}{a^3}-3ka+a^3V(T).\end{equation}We
see that the momentum $P_T$ appears linearly in the Hamiltonian.
This means that the parameter $T$ may be interpreted as a time
parameter when one deals with the quantum version of the model.
Considering the parameter $T$ as a clock parameter, the
Hamiltonian (\ref{G}) will be a time-dependent function. Such
Hamiltonian describes a system which exchanges energy with the
surrounding environment. However, in the case of cosmology where
the system under consideration is the whole universe, a
surrounding environment does not have any meaningful
interpretation. Therefore, such a Hamiltonian and the
corresponding time parameter do not seem to be suitable unless the
potential function is constant which for the sake of simplicity we
consider it to be zero. In this case the classical dynamics is
governed by the Hamiltonian equations, that is
\begin{eqnarray}\label{H}
\left\{
\begin{array}{ll}
\dot{a}=\{a,{\cal H}\}=-\frac{P_a}{6a},\\\\
\dot{P_a}=\{P_a,{\cal
H}\}=-\frac{P_a^2}{12a^2}+\frac{3P_T}{a^4}+3k,\\\\
\dot{T}=\{T,{\cal H}\}=\frac{1}{a^3},\\\\
\dot{P_T}=\{P_T,{\cal H}\}=0.
\end{array}
\right.
\end{eqnarray}Now, we would like to rewrite the above equations in
terms of the new time parameter $T$. By definition
$q'=\frac{dq}{dT}$ and with the help of the third equation of the
above system, the classical equations of motion can be rewritten
as follows
\begin{eqnarray}\label{I}
\left\{
\begin{array}{ll}
a'=-\frac{1}{6}a^2P_a,\\\\
P_a'=-\frac{1}{12}aP_a^2+\frac{3\Lambda}{a}+3ka^3,
\end{array}
\right.
\end{eqnarray}where we take $P_T=\Lambda=\mbox{const.}$ from the last equation of
(\ref{H}). We also have the constraint equation ${\cal H}=0$ from
which we obtain $P_a^2=\frac{12\Lambda}{a^2}-36ka^2$. Eliminating
$P_a$ between this equation and the first equation of (\ref{I})
results
\begin{equation}\label{J}
a'^2=\lambda^2a^2-ka^6,\end{equation}where $\lambda^2=\Lambda/3$.
In the following we present the solutions to this equation
according to the various values of the curvature index $k$.

$\bullet$ The flat universe:  $k=0$. In this case equation
(\ref{J}) admits the solutions
\begin{equation}\label{K}
a(T)=a_0e^{\pm\lambda T},\end{equation}where $a_0$ is an
integration constant and the positive (negative) sign corresponds
to an expanding (contracting) universe. For a positive sign in the
power of the exponential term, the evolutionary behavior of the
corresponding universe based on (\ref{K}) is like a de Sitter
universe, i.e., begins with a zero size at $t=-\infty$ and follows
the exponential law expansion at late time of cosmic evolution.
For a negative sign, on the other hand, the behavior is opposite.
The universe decreases its size from large values of scale factor
at $t=-\infty$ and tends to zero at the late time. In summary,
what we have shown is that a flat FRW model in the above gauge for
the time parameter has two separate classical solutions, one
describes an expanding universe while another represents a
contracting one. In figure \ref{fig3} (the left figure) these two
classical scale factors are plotted for typical values of the
parameters.

$\bullet$ The closed universe:  $k=1$. In this case equation
(\ref{J}) takes the form
\begin{equation}\label{L}
a'^2=\lambda^2a^2-a^6.\end{equation} It is clear from this
equation that the scale factor is restricted  to belong to the
interval $0\leq a \leq \sqrt{\lambda}$. Performing the integration
we get the following implicit relation between $T$ and $a(T)$
\begin{equation}\label{M}
\frac{a^2(T)}{\lambda^2+\lambda \sqrt{\lambda^2-a^4(T)}}={\cal
A}e^{\pm 2\lambda T},\end{equation}where ${\cal A}$ is an
integration constant. It is seen that the above mentioned
restriction on the values of the scale factor shows itself in the
expression under square root. Solving (\ref{M}) for $a(T)$, we get

\begin{equation}\label{M1}
a(T)=\frac{\sqrt{2{\cal A}}\lambda e^{\pm \lambda
T}}{\sqrt{1+\lambda^2 {\cal A}^2 e^{\pm 4\lambda T}}}.
\end{equation}
Again, we see that the classical closed model has also two
separate solutions (one for the positive sign in the power of the
exponential function and another for the negative sign) which we
have plotted them in figure \ref{fig5} (the left figure). As is
clear from the figure, in both of these models a contraction
period is followed by an expansion era. However, the turning point
in which the transition between two expanding and contracting
phases is occurred, is not the same for them.

$\bullet$ The open universe:  $k=-1$. In this case we have

\begin{equation}\label{N}
a'^2=\lambda^2a^2+a^6,\end{equation}where upon integration we
obtain
\begin{equation}\label{O}
\frac{a^2(T)}{\lambda^2+\lambda \sqrt{\lambda^2+a^4(T)}}={\cal
A}e^{\pm 2\lambda T},\end{equation}from which we have

\begin{equation}\label{P}
a(T)=\frac{\sqrt{2{\cal A}}\lambda e^{\pm \lambda
T}}{\sqrt{1-\lambda^2 {\cal A}^2 e^{\pm 4\lambda T}}}.
\end{equation}
These solutions are plotted in the left figure of figure
\ref{fig6}. As the figure shows, one branch of the classical
solutions begins with zero size at $t=-\infty$, increase the size
as time grows and tends to a singularity at which the scale factor
blows up. The another branch has the opposite behavior, i.e.,
begins with a singularity at which the scale factor has a large
size and then follows a contracting epoch and finally reaches the
zero size as $t\rightarrow +\infty$.
\section{Quantization of the model}
In this section we would like to see that how the classical
picture will be modified when one takes into account the quantum
mechanical considerations in the problem at hand. Before going to
do this, a remark about the canonical transformation which led us
to the Hamiltonian (\ref{G}) is in order. The canonical
transformation (\ref{F}) is applied to the classical Hamiltonian
(\ref{E}), resulting in Hamiltonian (\ref{G}) which we are going
to quantize. To make this acceptable, one should show that in the
quantum theory the two Hamiltonians are connected by some unitary
transformation, i.e. the transformation (\ref{F}) is also a
quantum canonical transformation. A quantum canonical
transformation is defined as a change of the phase-space variables
$\left(q,p\right)\rightarrow \left(q',p'\right)$ which preserves
the Dirac bracket, i.e.
$\left[q,p\right]=i=\left[q'(q,p),p'(q,p)\right]$. Such a
transformation is implemented by a unitary operator $C$ such that
$q'(q,p)=CqC^{-1}$ and $p'(q,p)=CpC^{-1}$ \cite{Arlen}. Under the
act of this unitary transformation state vectors will be
transformed as $\psi'=C\psi$ and any physical observable $A(q,p)$,
being an operator on the Hilbert space of states, transforms
according to $A'=CAC^{-1}$. Now, it is easy to see that if $C$ is
a unitary operator ($C^{\dag}=C^{-1}$) associated with some
quantum canonical transformation, then the Hermitian property does
not change after it transforms things from one coordinate system
to the other. Indeed, let $A=A^{\dag}$ be a Hermitian operator
associated to some physical observable. Under the act of a
canonical transformation $C$, we have
$A'^{\dag}=(CAC^{-1})^{\dag}=(C^{-1})^{\dag}A^{\dag}C^{\dag}=CAC^{-1}=A'$,
which means that $A'$ is also a Hermitian operator \footnote{At
this step a remark is in order about the self-adjointness problem
in the relation $T=\phi/P_{\phi}$. In general, the
self-adjointness of $\phi$ and $P_{\phi}$ does not lead to this
property for $T$, and with a non-self-adjoint operator the
construction of a self-adjoint Hamiltonian will be complicated.
However, note that what is explicitly appeared in the $T$-sector
of the Hamiltonian (\ref{G}) (with zero potential term) is $P_T$
(which is clearly a self-adjoint operator if $P_{\phi}$ is) and
not $T$. Also, in the $a$-sector of the Hamiltonian the
self-adjointness is encoded in the chosen factor ordering, see
equation (\ref{Q}) below. Therefore, we may claim that our theory
is based on a self-adjoint Hamiltonian. Since with the Hamiltonian
(\ref{G}), the quantization of the model yields a
Schr\"{o}dinger-like equation, in what follows, the role of the
variable $T$ looks like the time parameter in Schr\"{o}dinger
equation in terms of which the dynamics of the wave function and
other physical variables should be obtained.}. For our case it is
easy to see that the canonical relations
$\left[\phi,P_{\phi}\right]=i$ yield
$\left[T,P_T\right]=\left[\phi
P_{\phi}^{-1},\frac{1}{2}P_{\phi}^2\right]=i$. This means that the
transformation (\ref{F}) preserves the Dirac brackets and thus is
a quantum canonical transformation. Therefore, use of the
transformed Hamiltonian (\ref{G}) for quantization of the model is
quite reasonable.

Now, our starting point to quantize the model is to construct the
WDW equation $\hat{{\cal H}}\Psi(a,T)=0$, in which $\hat{{\cal
H}}$ is the Hamiltonian operator where its classical expression is
given by (\ref{G}) and $\Psi(a,T)$ is the wave function of the
universe. With the replacement $P_a\rightarrow
-i\frac{\partial}{\partial a}$ and similarly for $P_T$ and
choosing the potential to be zero, the WDW equation reads
\begin{equation}\label{Q}
\left(a^2\frac{\partial^2}{\partial
a^2}+sa\frac{\partial}{\partial a}-12i\frac{\partial}{\partial
T}-36ka^4\right)\Psi(a,T)=0,\end{equation}where the parameter $s$
represents the ambiguity in the ordering of factors $a$ and $P_a$
in the first term of (\ref{G}). This equation takes the form of a
Schr\"{o}dinger equation and we separate its variables as
\begin{equation}\label{R}
\Psi(a,T)=e^{iET}\psi(a),\end{equation}leading to
\begin{equation}\label{S}
\left[a^2\frac{d^2}{da^2}+sa\frac{d}{da}+12(E-3ka^4)\right]\psi(a)=0,\end{equation}where
$E$ is a separation constant. In the following, as in the
classical model, we shall deal with the solution to this equation
in $k=\pm 1, 0$ cases separately.

$\bullet$ The flat universe:  $k=0$. In this case equation
(\ref{S}) has the following solutions
\begin{equation}\label{T}
\psi_E(a)=c_1\sin \left(2\sqrt{3E}\ln a\right)+c_2\cos
\left(2\sqrt{3E}\ln a\right),\end{equation}where $c_1$ and $c_2$
are the integration constants and we have chosen $s=1$. If,
without losing the general character of the solutions, we set
$c_2=0$ the eigenfunctions of the SWD equation can be written as

\begin{equation}\label{U}
\Psi_E(a,T)=e^{iET}\sin \left(2\sqrt{3E}\ln
a\right).\end{equation}We may now write the general solution to
the SWD equation as a superposition of its eigenfunctions, that is
\begin{equation}\label{V}
\Psi(a,T)=\int_0^\infty A(E)\Psi_E(a,T)dE,\end{equation}where
$A(E)$ is a suitable weight function to construct the wave
packets. By using the equality
\begin{equation}\label{X}
\int_0^\infty e^{-\gamma x}\sin\sqrt{mx}dx=\frac{\sqrt{\pi
m}}{2\gamma^{3/2}}e^{-(m/4\gamma)},\end{equation}we can evaluate
the integral over $E$ in (\ref{X}) and simple analytical
expression for this integral is found if we choose the function
$A(E)$ to be a quasi-Gaussian weight factor $A(E)=e^{-\gamma E}$
($\gamma$ is an arbitrary positive constant), which results in
\begin{equation}\label{Y}
\Psi(a,T)=\int_0^\infty e^{-\gamma E}e^{iET}\sin
\left(2\sqrt{3E}\ln a\right)dE.\end{equation}Using of the relation
(\ref{X}) leads to the following expression for the wave function
\begin{equation}\label{Z}
\Psi(a,T)={\cal N}\frac{\ln a}{(\gamma-iT)^{3/2}}\exp
\left(-\frac{3\ln^2 a}{\gamma-iT}\right),\end{equation}where
${\cal N}$ is a numerical factor.

\begin{figure}
\begin{tabular}{c}\epsfig{figure=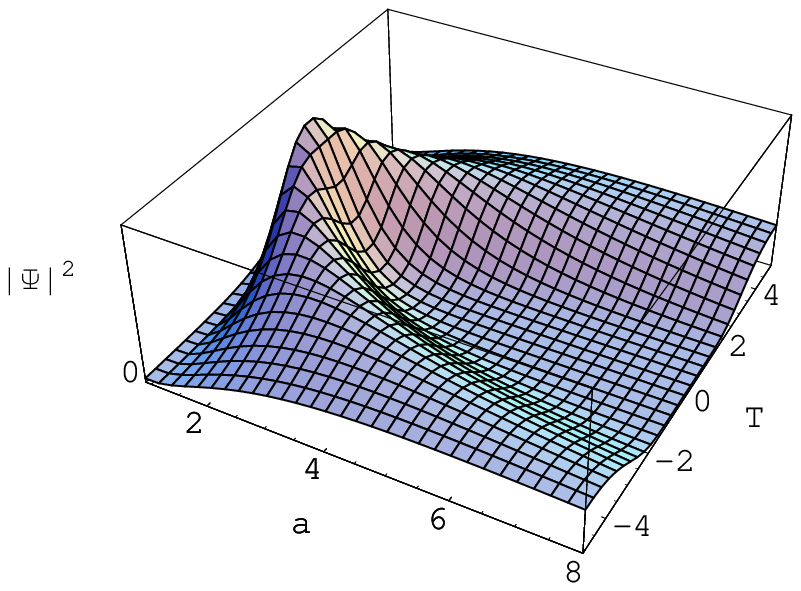,width=5cm}
\hspace{1.5cm} \epsfig{figure=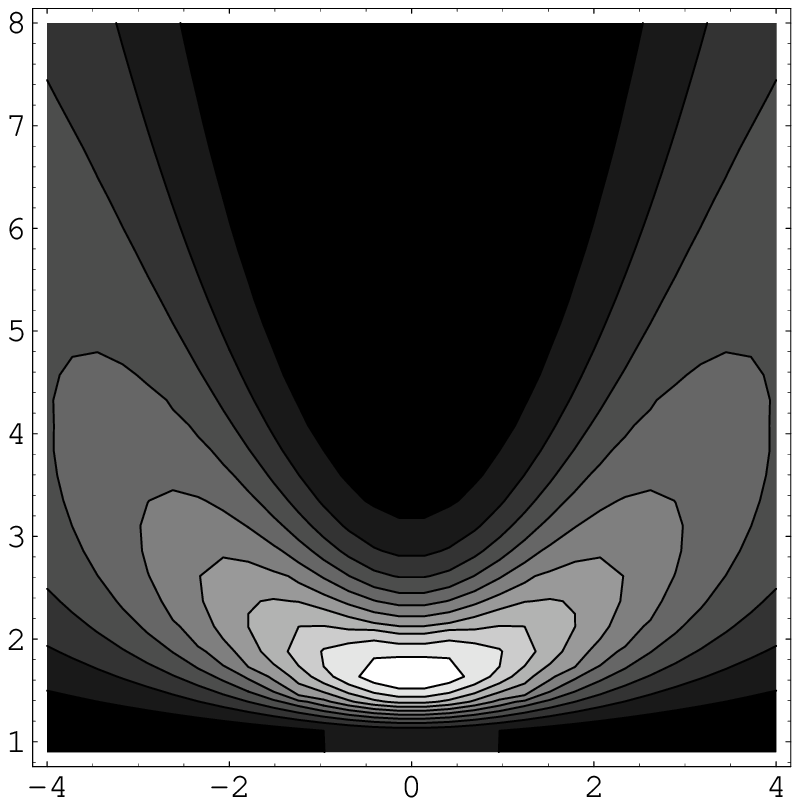,width=4cm}
\end{tabular}
\caption{\footnotesize  The figures show the square of the wave
function (\ref{Z}) and its contourplot.}\label{fig2}
\end{figure}
In figure \ref{fig2} we have plotted the square of the wave
function and its contourplot for typical numerical values of the
parameters. As this figure shows, at $T=0$, the wave function has
a dominant peak in the vicinity of some non-zero value of $a$.
This means that the wave function predicts the emergence of the
universe from a state corresponding to its dominant peak. As time
progresses, the wave packet begins to propagate in the
$a$-direction, its width becoming wider and its peak moving with a
group velocity towards the greater values of scale factor. The
wave packet disperses as time passes, the minimum width being
attained at $T=0$. As in the case of the free particle in quantum
mechanics, the more localized the initial state at $T=0$, the more
rapidly the wave packet disperses. Therefore, the quantum effects
make themselves felt only for small enough $T$ corresponding to
small $a$, as expected and the wave function predicts that the
universe will assume states with larger $a$ in its late time
evolution. Now, having the above expression for the wave function
of the universe, we are going to deal with the question of the
recovery of the classical solutions, in other words, how can the
Wheeler-DeWitt wave functions predict a classical universe. In
this approach, as we have done above, one usually constructs a
coherent wave packet with good asymptotic behavior in the
minisuperspace, peaking in the vicinity of the classical
trajectory. On the other hand, to show the correlations between
classical and quantum pattern, following the many-worlds
interpretation of quantum mechanics \cite{tip}, one may calculate
the time dependence of the expectation value of a dynamical
variable $q$ as
\begin{equation}\label{AB}
<q>(t)=\frac{<\Psi|q|\Psi>}{<\Psi|\Psi>}.\end{equation} Following
this approach, we may write the expectation value for the scale
factor as
\begin{equation}\label{AC}
<a>(T)=\frac{\int_0^\infty \Psi^*(a,T)a\Psi(a,T)da}{\int_0^\infty
\Psi^*(a,T)\Psi(a,T)da},\end{equation}which yields
\begin{equation}\label{AD}
<a>(T)\sim
\frac{T^2+\gamma^2+3\gamma}{T^2+\gamma^2+12\gamma}e^{\frac{T^2+\gamma^2}{8\gamma}}.\end{equation}
It is important to classify the nature of the quantum model as
concerns the presence or absence of singularities. For the wave
function (\ref{Z}), the expectation value (\ref{AD}) of scale
factor never vanishes, showing that these states are nonsingular.
Indeed, the expression (\ref{AD}) represents a bouncing universe
with no singularity where its late time behavior coincides to the
late time behavior of the classical solution. This means that the
quantum structure which we have constructed has a good correlation
with its classical counterpart.

\begin{figure}
\begin{tabular}{c}\epsfig{figure=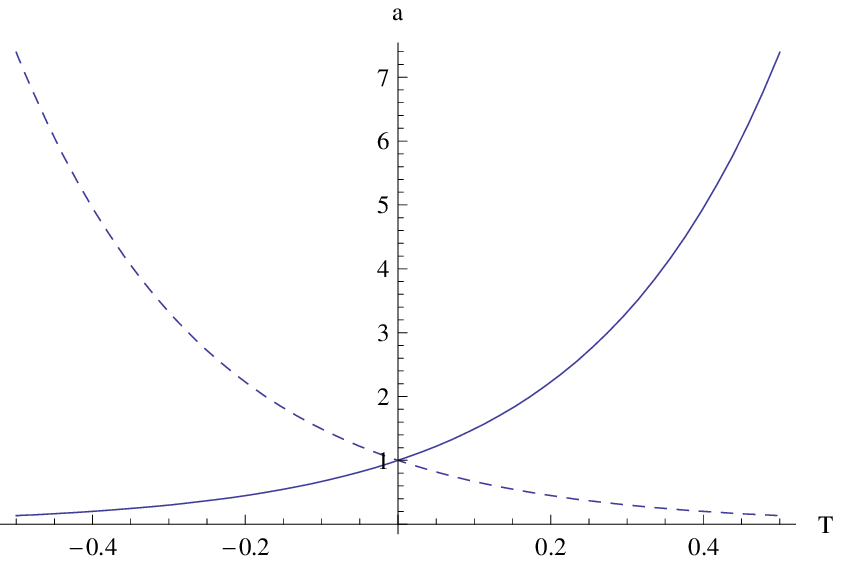,width=4cm}
\hspace{1.5cm} \epsfig{figure=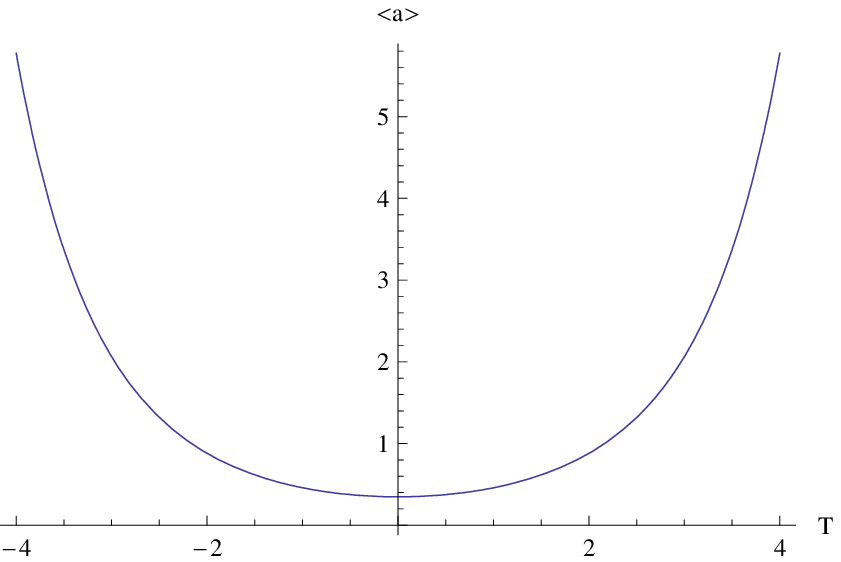,width=4cm}\hspace{1.5cm}
\epsfig{figure=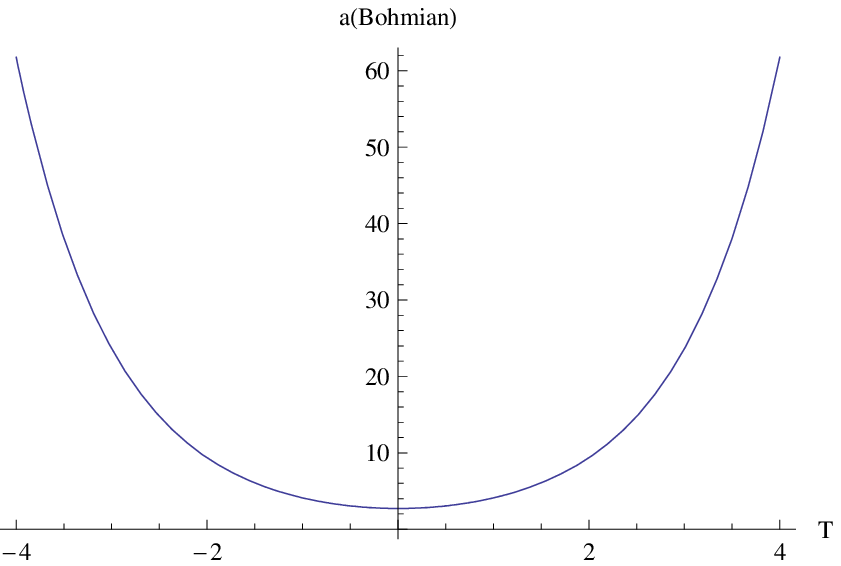,width=4cm}
\end{tabular}
\caption{\footnotesize  From left to right: The classical scale
factors (\ref{K}), the expectation value (\ref{AD}) and the
Bohmian value (\ref{AH}).}\label{fig3}
\end{figure}

The issue of the correlation between classical and quantum schemes
may be addressed from another point of view. It is known that the
results obtained by using the many-world interpretation agree with
those that can be obtained using the ontological interpretation of
quantum mechanics \cite{bohm}. In Bohmian interpretation, the wave
function is written as $\Psi(a,T)=\Omega(a,T)e^{iS(a,T)}$, where
$\Omega(a,T)$ and $S(a,T)$ are real functions, which using the
expression (\ref{Z}), a simple algebra gives them as
\begin{equation}\label{AE}
\Omega(a,T)={\cal N}\left(\gamma^2+T^2\right)^{-3/4}
\exp\left(-3\gamma \frac{\ln^2 a}{\gamma^2+T^2}\right)\ln
a,\end{equation}
\begin{equation}\label{AF}
S(a,T)=\frac{3}{2}\arctan \frac{T}{\gamma}-\frac{3T \ln^2
a}{\gamma^2+T^2}.\end{equation}In this interpretation the
classical trajectories, which determine the behavior of the scale
facto is given by $P_a=\frac{\partial S}{\partial a}$, where by
means of the relation (\ref{D}) reads
\begin{equation}\label{AG}
\frac{1}{a}\frac{da}{dT}=\frac{T}{\gamma^2+T^2}\ln
a,\end{equation}which, after integration we get
\begin{equation}\label{AH}
a(T)=a_0e^{\sqrt{\gamma^2+T^2}}.\end{equation}This solution has
the same behavior as the expectation values computed in (\ref{AD})
and like that is free of singularity. figure \ref{fig3} shows the
behavior of the classical scale factor (\ref{K}), the quantum
mechanical expectation value of the scale factor (\ref{AD}) and
its Bohmian version (\ref{AH}) versus time for some typical
numerical values of the parameters. We see that instead of two
separate classical solutions, the quantum models predict a
bouncing universe in which the universe decreases its size,
reaches a minimum and then expands forever. The origin of the
singularity avoidance may be understood by the existence of the
quantum potential which corrects the classical equations of
motion. According to the Bohm-de Broglie interpretation of quantum
mechanics and also its usage in quantum cosmology, upon using the
polar form of the wave function
$\Psi(a,T)=\Omega(a,T)e^{iS(a,T)}$, in the corresponding wave
equation, we arrive at the modified Hamilton-Jacobi equation as
\begin{equation}\label{AI}
{\cal H}\left(q_i,P_i=\frac{\partial S}{\partial q_i}\right)+{\cal
Q}=0,\end{equation}where $P_i$ are the momentum conjugate to the
dynamical variables $q_i$ and ${\cal Q}$ is the quantum potential.
For the wave equation (\ref{Q}) this procedure leads
\begin{equation}\label{AJ}
-\frac{1}{12a}\left(\frac{\partial S}{\partial
a}\right)^2+\frac{1}{a^3}\left(\frac{\partial S}{\partial
T}\right)-3ka+{\cal Q}=0,\end{equation}in which the quantum
potential is defined as
\begin{equation}\label{AL}
{\cal Q}=\frac{1}{12 a \Omega}\frac{\partial^2\Omega}{\partial
a^2}+\frac{1}{12a^2\Omega}\frac{\partial \Omega}{\partial
a}.\end{equation}Now, inserting the relation (\ref{AH}) in
(\ref{AL}), we can find the quantum potential in terms of the
scale factor as
\begin{equation}\label{AM}
{\cal Q}\sim \frac{1}{a^3 \ln^2(\frac{a}{a_0})}.\end{equation}As
this relation shows, the potential goes to zero for the large
values of the scale factor. This behavior is expected, since in
this regime the quantum effects can be neglected and the universe
evolves classically. On the other hand, for the small values of
the scale factor the potential takes a large magnitude and the
quantum mechanical considerations come into the scenario. This is
where the quantum potential can produce a huge repulsive force,
which may be interpreted as being responsible of the avoidance of
singularity.

$\bullet$ The closed universe:  $k=1$. Equation (\ref{S}) in this
case admits the following solutions in terms of the modified
Bessel functions
\begin{equation}\label{AN}
\psi_E(a)=c_1K_{i\sqrt{3E}}\left(3a^2\right)+c_2I_{i\sqrt{3E}}\left(3a^2\right).\end{equation}We
impose the boundary condition $\lim_{x\rightarrow\infty}
\psi(x)=0$ on the wave function which restricts us to consider the
modified Bessel function $K$ as solution. Thus the eigenfunctions
of the SWD equation take the form
\begin{equation}\label{AO}
\Psi_E(a,T)=e^{iET}K_{i\sqrt{3E}}\left(3a^2\right),\end{equation}which
their superposition gives the general solution as
\begin{equation}\label{AP}
\Psi(a,T)=\int_0^\infty \frac{2}{3}\nu C(\nu)e^{\frac{1}{3}i\nu^2
T}K_{i\nu}\left(3a^2\right)d\nu,\end{equation}where
$\nu=\sqrt{3E}$. Since the function $K_{i\nu}(z)$ decreases when
$\nu$ grows, thus the small $\nu$'s have dominate contribution to
the above integral. This allows us to use the approximation
$\nu^2/3\sim \nu$. Therefore, with the choice of the weight factor
as $C(\nu)=\frac{3}{2}e^{\gamma \nu}$ we get
\begin{equation}\label{AR}
\Psi(a,T)=\int_0^\infty \nu
e^{i(T-i\gamma)\nu}K_{i\nu}\left(3a^2\right)d\nu.\end{equation}Now,
by using the equalities \cite{Handbook}
\begin{equation}\label{AS}
\int_0^\infty \cos (\beta
\nu)K_{i\nu}(\alpha)d\nu=\frac{\pi}{2}e^{-\alpha \cosh
\beta},\hspace{0.3cm}\int_0^\infty \nu \sin (\beta
\nu)K_{i\nu}(\alpha)d\nu=\frac{\pi}{2}\alpha \sinh \beta
e^{-\alpha \cosh \beta},\end{equation}for $\alpha>0$ and
$|\mbox{Im} \beta|<\frac{\pi}{2}$, we can evaluate the integral in
(\ref{AR}) to achieve the following analytic form for the wave
function
\begin{equation}\label{AT}
\Psi(a,T)={\cal N}\left[1+3a^2\sin
\left(\gamma+iT\right)\right]e^{-3a^2\cos\left(\gamma+iT\right)},\end{equation}where
again ${\cal N}$ is a numerical factor and $\gamma$ should be
chosen as $|\gamma|<\frac{\pi}{2}$.
\begin{figure}
\begin{tabular}{c}\epsfig{figure=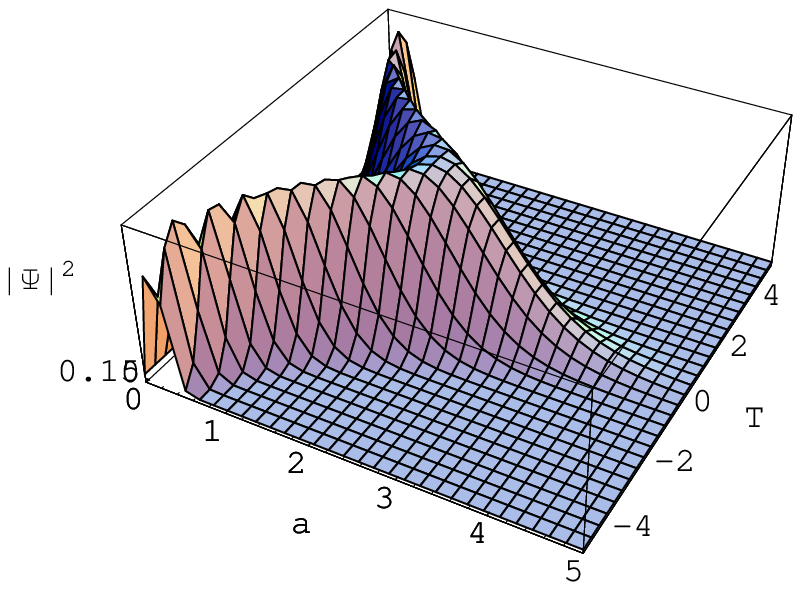,width=5cm}
\hspace{1.5cm} \epsfig{figure=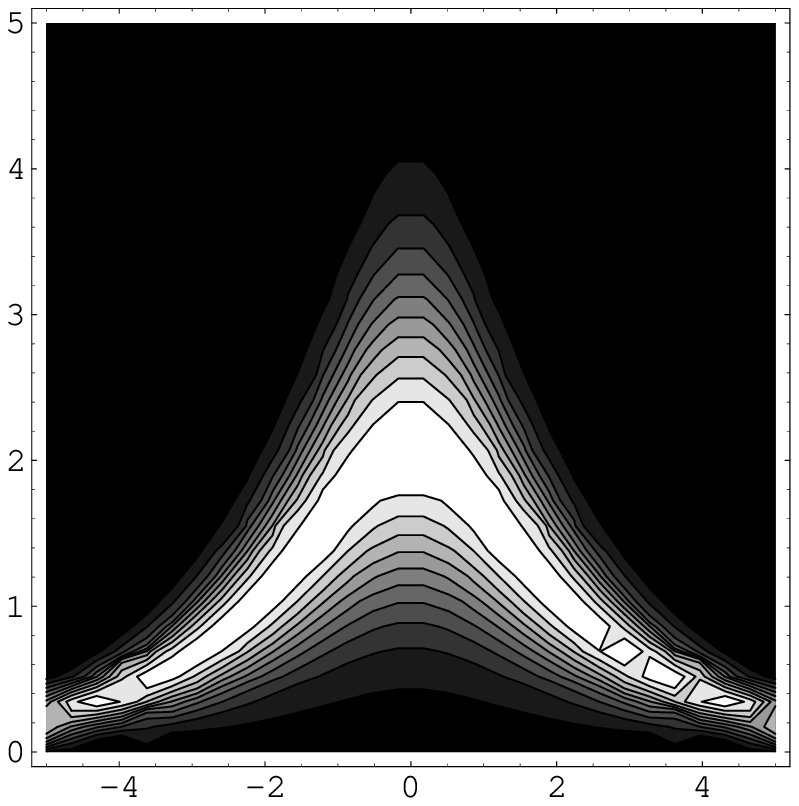,width=4cm}
\end{tabular}
\caption{\footnotesize  The figures show the square of the wave
function (\ref{AT}) and its contourplot.}\label{fig4}
\end{figure}
In figure \ref{fig4} the wave function and its contourplot is
plotted. This figure shows the universe expands and after reaches
a maximum recollapses. It is clear that the two phases of cosmic
evolution are connected to each other without meeting any kind of
singularity. Now, with the help of the definition (\ref{AB}) we
can evaluate the expectation value of the scale factor when the
corresponding quantum universe is in the state (\ref{AT}). This
calculation results
\begin{eqnarray}\label{AU}
&&<a>(T)=\cosh^{-1/2}T \times \nonumber\\&&
\hspace{-2.2cm}\frac{8\left(e^T+e^{3T}+e^{5T}\right)+\left(1+e^{2T}\right)^3\cos
\gamma+4e^{T}\left(1+e^{4T}\right)\cos 2\gamma-\left(1+
3e^{2T}+3e^{4T}+e^{6T}\right)\cos
3\gamma}{11e^T+16e^{3T}+11e^{5T}+\left(1+e^{2T}\right)^3\cos
\gamma+2\left(4+5e^{2T}+4e^{4T}\right)\cos
2\gamma-\left(1+3e^{2T}+3e^{4T}+e^{6T}\right)\cos 3\gamma},
\end{eqnarray}
where is plotted in figure \ref{fig5}. Also, to extract the
Bohmian trajectories of this model we may write the wave function
(\ref{AT}) as $\Psi(a,T)=\Omega(a,T)e^{iS(a,T)}$ in which
\begin{equation}\label{AV}
\Omega(a,T)=\left[1+9a^4\left(\sin^2\gamma \cosh ^2T+\cos^2\gamma
\sinh^2 T\right)+6a^2\sin^2\gamma
\cosh^2T\right]^{1/2}e^{-3a^2\cos \gamma \cosh
T},\end{equation}and
\begin{equation}\label{AX}
S(a,T)=3a^2\sin \gamma \sinh T+\arctan \frac{3a^2\cos \gamma \sinh
T}{1+3a^2\sin \gamma \cosh T}.\end{equation}Substitution of the
relation (\ref{AX}) in $P_a=\frac{\partial S}{\partial a}$ and
using of the definition $P_a$ in (\ref{D}), we obtain
\begin{equation}\label{AY}
\frac{da}{dT}=-\frac{a^3\sinh T\left[\cos \gamma+\sin
\gamma-\frac{9}{2}a^4\left(\cos 2\gamma-\cosh 2T\right)\sin
\gamma+6a^2\cosh T\sin^2\gamma\right]}{1-\frac{9}{2}a^4\left(\cos
2\gamma-\cosh 2T\right)+6a^2\cosh T \sin
\gamma}.\end{equation}This equation does not seem to have
analytical solution. However, in figure \ref{fig5}, employing
numerical methods, we have shown the qualitative behavior of
$a(T)$ for typical values of the parameters and initial
conditions. As is clear from the figure, the scale factor repeats
its expansion and contraction behavior as its expectation value
also shown in this figure. We see that the quantization of the
closed FRW cosmology predicts an evolution pattern for the
corresponding universe in completely agreement with its classical
dynamics which means that there is a good correlation between the
classical and quantum models in this case as well.

\begin{figure}
\begin{tabular}{c}\epsfig{figure=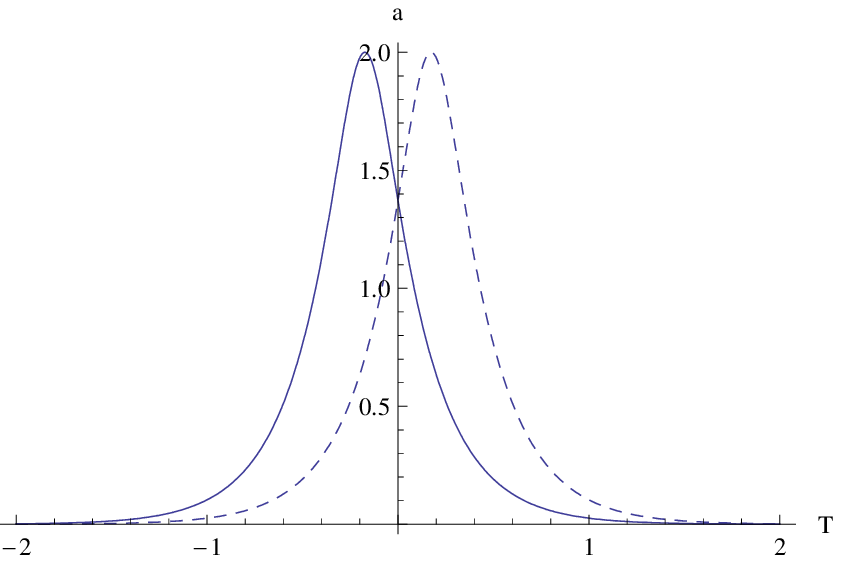,width=4cm}\hspace{1.5cm}\epsfig{figure=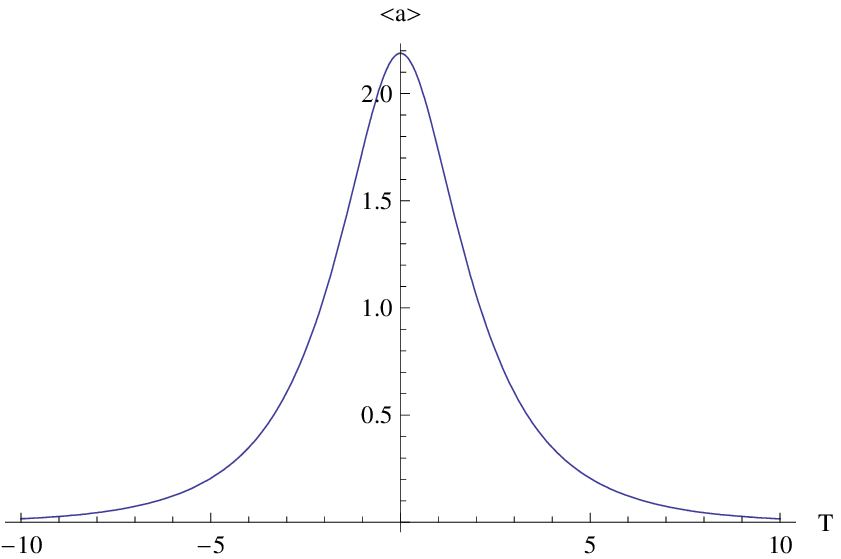,width=4cm}
\hspace{1.5cm} \epsfig{figure=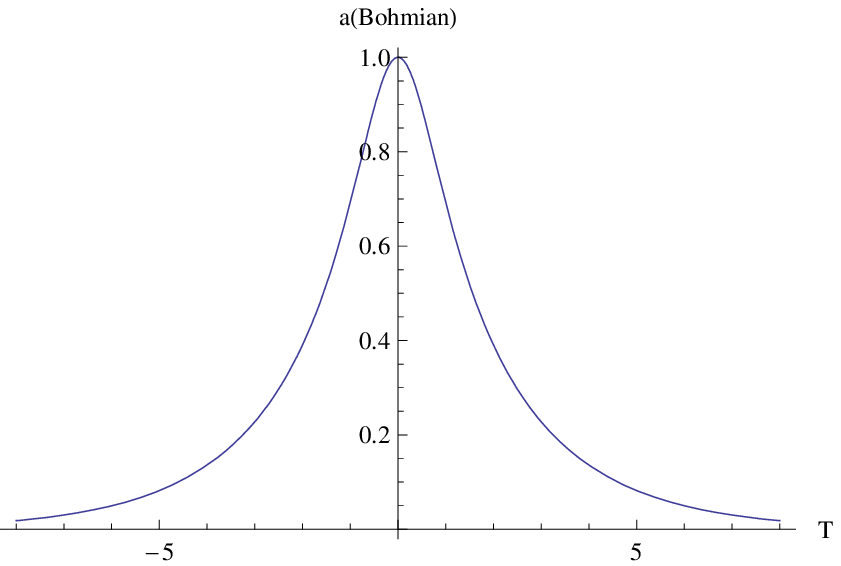,width=4cm}
\end{tabular}
\caption{\footnotesize  From left to right: The classical scale
factors (\ref{M1}), the expectation value (\ref{AU}) and the
Bohmian trajectory obtained from (\ref{AY}).}\label{fig5}
\end{figure}

\begin{figure}
\begin{tabular}{c}\epsfig{figure=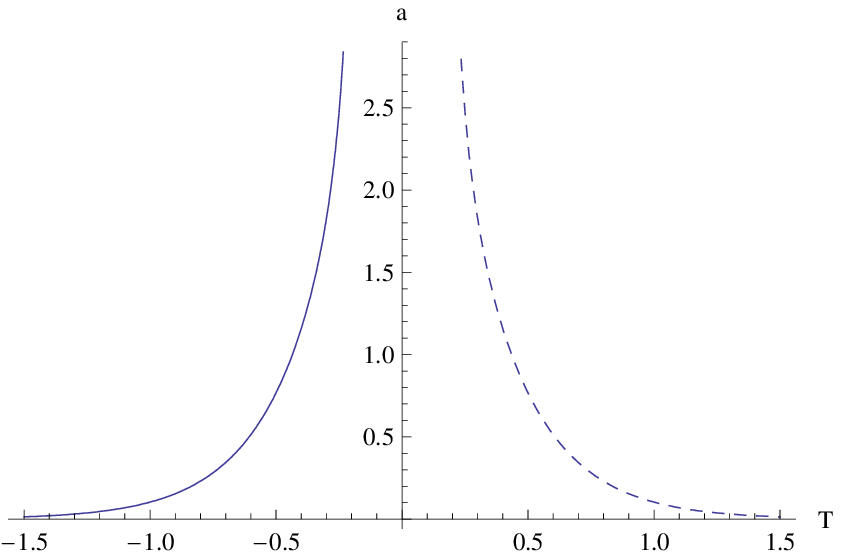,width=4cm}\hspace{1.5cm}\epsfig{figure=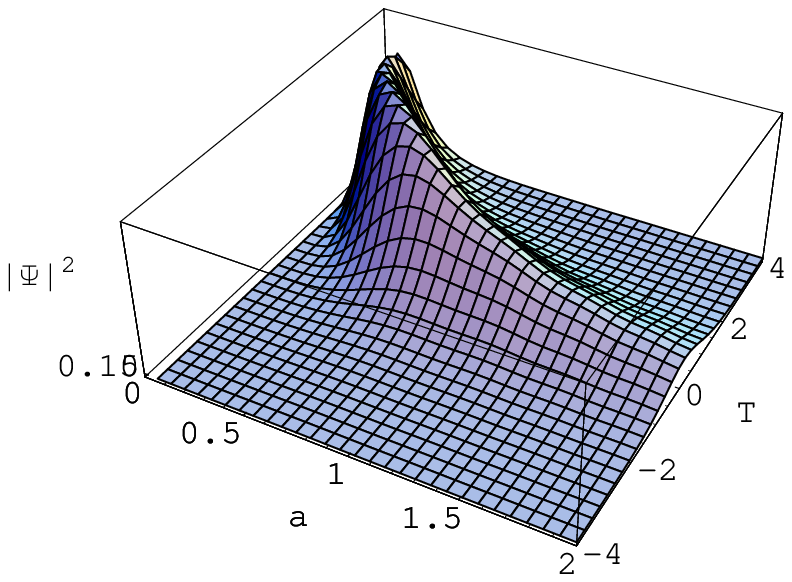,width=4cm}
\hspace{1.5cm} \epsfig{figure=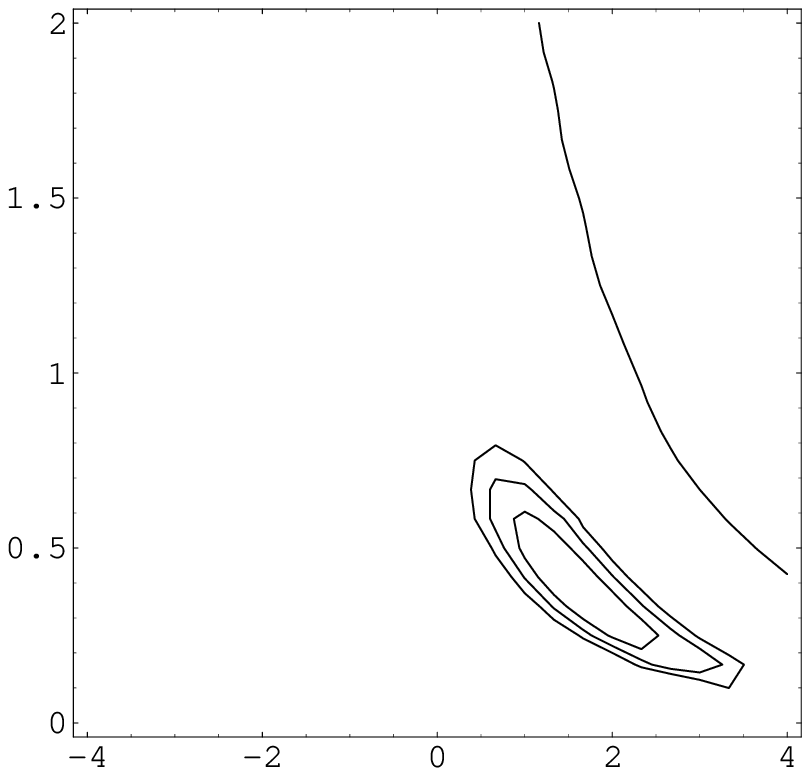,width=4cm}
\end{tabular}
\caption{\footnotesize From left to right: The classical scale
factors (\ref{P}), the square of the wave function (\ref{BC}) and
its  contourplot.}\label{fig6}
\end{figure}

$\bullet$ The open universe:  $k=-1$. The solutions to the
equation (\ref{S}) with $k=-1$ read as
\begin{equation}\label{AZ}
\psi_E(a)=c_1J_{i\sqrt{3E}}(3a^2)+c_2Y_{i\sqrt{3E}}(3a^2),\end{equation}where
$J$ and $Y$ are the Bessel functions. Removing the function $Y$
from the solutions because of applying the same boundary condition
as in the closed case, we are led to the following eigenfunctions
\begin{equation}\label{BA}
\Psi_E(a,T)=e^{iET}J_{i\sqrt{3E}}(3a^2). \end{equation}We may now
write the general solutions to the SWD equation as a superposition
of the above eigenfunctions
\begin{equation}\label{BC}
\Psi(a,T)=\int_0^\infty
C(E)e^{iET}J_{i\sqrt{3E}}(3a^2)dE.\end{equation}Unfortunately,
unlike the previous two cases, it is not possible to find an
analytical closed form for the wave function in the case of the
open universe. Therefore, we should rely on the numerical methods
by using of the shifted Gaussian weight function
$C(E)=e^{-\gamma(E-\sigma)^2}$. In figure \ref{fig6} we have
plotted the square of the wave function and its contourplot. The
discussions on the comparison between quantum cosmological
solutions and their corresponding form from the classical
formalism are the same as previous models, namely the flat and
closed models. Similar discussion as above would be applicable to
this case as well.
\section{Summary}
In this letter we took a look at the problem of time in the
framework of a scalar field cosmology. To do this, we have
examined a canonical transformation $(\phi,P_{\phi})\rightarrow
(T,P_T)$ on the scalar field sector of the action which turns out
to correspond to a specific choice of time parameter. In terms of
this time parameter, we have solved the classical field equations
in the cases of the flat, closed and open FRW universe and
obtained the analytical expressions for the corresponding scale
factors. As for the quantum version of these models, the use of
the above mentioned canonical transformation allowed us to obtain
a SWD equation in which the variable $T$ plays the role of time
parameter. In the cases of the flat and closed FRW models, we
obtained exact solutions of the SWD equation and for the open
model the wave function is represented numerically. The wave
functions of the corresponding universes consist of some branches
where each may be interpreted as part of the classical
trajectories. We saw that since the peaks of the wave functions
follow the classical trajectories, there seems to be good
correlation between the corresponding classical and quantum
cosmology.

One of the main features of the quantum solutions in our presented
model is that the singular region of the classical cosmologies is
replaced by a bouncing period. In such a behavior which is shown
in figures \ref{fig2}-\ref{fig6} the scale factor bounces (falls)
from the contraction (expansion) to its expansion (contraction)
eras. In addition to singularity avoidance, the appearance of
bounce in the quantum model is also interesting in its nature due
to prediction of a minimal size for the corresponding universe. We
know the idea of existence of a minimal length in nature is
supported by almost all candidates of quantum gravity. In this
sense our results are qualitatively comparable with other works in
which different quantization schemes are considered. We may name
some of such works, for instance, as:

$\bullet$ In \cite {Cor}, using the polymer quantization method of
a $k=0$ FRW model, the volume of the corresponding universe is
obtained as $\sqrt{\gamma^2 \lambda^2+9t^2}$, where $\gamma$ is
the Barbero-Immirzi parameter and $\lambda$ is a parameter
associated to the fundamental granularity of quantum geometry.

$\bullet$ In \cite{Bab}, using a gauge-fixed Lagrangian to
quantize a perfect fluid FRW cosmology, a bouncing scale factor
$(\gamma^2+t^2)^{\frac{1}{3(\omega+1)}}$, for ordinary perfect
fluid and a falling one
$[\gamma^2+(t_0-t)^2]^{-\frac{1}{3(\omega+1)}}$ for phantom
perfect fluid is obtained, where $\gamma$ is an arbitrary positive
constant and $\omega$ is the equation of state parameter of the
perfect fluid.

$\bullet$ In \cite{Fab}, a perfect fluid FRW cosmology is
quantized with the use of the Schutz' representation of the
perfect fluid which is shown may lead to an identification to a
time parameter. The bouncing expression such as
$[\frac{9(1-\omega)^4}{\gamma^2}t^2+1]^{\frac{1}{3(1-\omega)}}$ is
obtained for the scale factor in this model.

$\bullet$ In \cite{Nov} several bouncing solutions of various
cosmological models are investigated and the mechanisms behind the
bounce are discussed.\vspace{5mm}\newline \noindent {\bf
Acknowledgements}\vspace{2mm}\noindent\newline The author is
grateful to the research council of IAU, Chalous Branch for
financial support.

\end{document}